\documentclass{mnras}

\usepackage{natbib}
\usepackage{graphicx}
\usepackage{epstopdf}
\usepackage[T1]{fontenc}
\usepackage{aecompl}
\usepackage{mnras-bst}
\usepackage{hyperref}
\bibpunct{(}{)}{;}{a}{}{,}

\usepackage[T1]{fontenc}
\usepackage{ae,aecompl}

\usepackage{amsmath}

\newcommand\0{\phantom{0}}

\title{The wavelength dependence of interstellar polarization in the Local Hot Bubble}

\author[D. V. Cotton et al.]{Daniel V. Cotton$^{1,2}$\thanks{E-mail: d.cotton@unsw.edu.au}, Jonathan P. Marshall$^{3,4}$, Priscilla C. Frisch$^{5}$,
\newauthor Lucyna Kedziora-Chudzer$^{1,2}$, Jeremy Bailey$^{1,2}$, Kimberly Bott$^{6,7}$,
\newauthor Duncan J. Wright$^{1,4,8}$, Mark C. Wyatt$^{9}$ and Grant M. Kennedy$^{10}$.
\\
\\
$^{1}$School of Physics, UNSW Australia, High Street, Kensington, NSW 2052, Australia. \\
$^{2}$Australian Centre for Astrobiology, UNSW Sydney, NSW 2052, Australia. \\
$^{3}$Academia Sinica, Institute of Astronomy and Astrophysics, 11F Astronomy-Mathematics Building, NTU/AS campus, No. 1, \\
$^{\0}$Section 4, Roosevelt Rd., Taipei 10617, Taiwan. \\
$^{4}$Centre for Astrophysics, University of Southern Queensland, Toowoomba, Qld. 4350, Australia. \\ 
$^{5}$Department of Astronomy and Astrophysics, University of Chicago, Chicago, IL 60637, USA. \\
$^{6}$NExSS Virtual Planetary Laboratory, Seattle, WA 98195, USA. \\
$^{7}$Astronomy Department, University of Washington, Box 351580, Seattle, WA 98195, USA. \\
$^{8}$Australian Astronomical Observatory (AAO), P.O. Box 915, North Ryde, NSW 1670, Australia. \\
$^{9}$Institute of Astronomy, University of Cambridge, Madingley Road, Cambridge, CB3 0HA, UK. \\
$^{10}$Department of Physics, University of Warwick, Gibbet Hill Road, Coventry, CV4 7AL, UK.\\
}

\date{Accepted XXX. Received YYY; in original form ZZZ}

\pubyear{2018}

\begin{document}
\label{firstpage}
\pagerange{\pageref{firstpage}--\pageref{lastpage}}
\maketitle

\begin{abstract}
The properties of dust in the interstellar medium (ISM) nearest the Sun are poorly understood because the low column densities of dust toward nearby stars induce little photometric reddening, rendering the grains largely undetectable. Stellar polarimetry offers one pathway to deducing the properties of this diffuse material. Here we present multi-wavelength aperture polarimetry measurements of seven bright stars chosen to probe interstellar polarization near the edge of the Local Hot Bubble (LHB) - an amorphous region of relatively low density interstellar gas and dust extending $\sim$70--150~pc from the Sun. The measurements were taken using the HIgh Precision Polarimetric Instrument (HIPPI) on the 3.9-m Anglo-Australian Telescope. HIPPI is an aperture stellar polarimeter with a demonstrated sensitivity of 4.3~parts-per-million (ppm). Of the stars observed two are polarized to a much greater degree than the others; they have a wavelength of maximum polarization ($\lambda_{max}$) of $\sim$550~$\pm$~20~nm -- similar to that of stars beyond the LHB -- and we conclude that they are in the wall of the LHB. The remaining five stars have polarizations of $\sim$70 to 160~ppm, of these four have a much bluer $\lambda_{max}$, $\sim$350~$\pm$~50~nm. Bluer values of $\lambda_{max}$ may indicate grains shocked during the evolution of the Loop I Superbubble. The remaining star, HD~4150 is not well fit by a Serkowski curve, and may be intrinsically polarized.
\end{abstract}




\section{Introduction}

The light reaching us from all stars is polarized by aligned dust grains in the interstellar medium (ISM). Alignment is primarily caused by the interstellar magnetic field (ISMF). The magnitude of interstellar polarization depends on the degree of alignment of the local magnetic field (magnetic turbulence), the angle between the field and the line of sight, and the degree the field varies along the line of sight \citep{draine03,jones15}. Polarimetric studies of sufficiently intrinsically unpolarized stars probe this interstellar polarization to reveal the structure of the ISMF as well as the properties and history of dust in the ISM \citep{clarke10,frisch15,jones15}.
 
The wavelength dependence of interstellar (linear) polarization is given by the empirically determined Serkowski Law \citep{serkowski75} as: \begin{equation}\label{eq:Serk} \frac{p(\lambda)}{p_{max}}=exp\left[-Kln^2\left (\frac{\lambda_{max}}{\lambda} \right )  \right ] ,\end{equation} where $p(\lambda)$ is the polarization at wavelength $\lambda$, $p_{max}$ is the maximum polarization occurring at wavelength $\lambda_{max}$. The dimensionless constant $K$ describes the inverse width of the polarization curve peaked around $\lambda_{max}$; \citet{serkowski75} gave its value as 1.15. Later authors have refined this relation,  with \citet{wilking80} being the first to describe $K$ in terms of a linear function of $\lambda_{max}$ -- consequently the relationship is sometimes called the Serkowski Law, or the Serkowski-Wilking Law. Using this form \citet{whittet92} give $K$ as: \begin{equation}\label{eq:Whit} K=(0.01\pm0.05)+(1.66\pm0.09)\lambda_{max},\end{equation} (where $\lambda_{max}$ is given in $\mu$m) which they indicate is applicable for the wavelength range 360~nm to 2000~nm. A typical value for $\lambda_{max}$ within our galaxy is 550~nm \citep{serkowski75}, but a wide range of extremes have been reported, e.g. 360~nm to 890~nm \citep{wilking82}, and smaller values are more common in other galaxies \citep{jones15}. However, all previous work corresponds to regions beyond the Local Hot Bubble (LHB).

The wavelength dependence of interstellar polarization in the ISM reveals the size of the dust grains within it. The reddest values of $\lambda_{max}$, for instance, are associated with dusty nebulae and larger grain sizes \citep{whittet92,clarke10}. As a consequence of scattering processes, the wavelength of maximum polarization, $\lambda_{max}$, is proportional to $2\pi a$, where $a$ corresponds to the radius of the dominant dust grains \citep{draine95}. It has been shown that carbonaceous grains and those smaller than 100~nm tend not to contribute to interstellar polarization, so that the dominant grains tend to be the larger silicate (and ice) grains \citep{jones15}. Though \citet{papoular18} has recently proposed an alternate interpretation where $\lambda_{max}$ depends not on grain size but on the relative composition of enstantite and fosterite in the silicon grains, the current picture is long established \citep{clarke10,jones15}.

The LHB is a region largely devoid of gas that extends out to $\sim$70--150~pc from the Sun. The dust density in the LHB is also low, though the exact correspondence between the extent of the gas and dust bubbles has not been established. Owing to the low dust density, here interstellar polarization is very low, and its properties poorly understood. Outside the LHB interstellar polarization typically\footnote{Such measurements are complicated since the distribution of gas and dust within the ISM is often filamentary (e.g. \citealp{arzoumanian11}), meaning different trends may manifest on adjacent sight lines.} increases with distance at a rate of tens of ppm/pc \citep{behr59}, whereas inside the LHB this figure is 0.2--2~ppm/pc \citep{bailey10,cotton16a}. Until recently neither reddening nor polarimetric measurements were possible at the required precision to measure interstellar dust within the LHB. With the advent of modern high-precision polarimeters it is now possible to map the ISM where column densities are low, N(H$^0$)~$<$~10$^{18.5}$ \textit{\citep{frisch15}}. A number of recent high precision polarimetric works have made small surveys of parts of the LHB \citep{bailey10,cotton16a,cotton16b,frisch16,cotton17a}, and the regions just beyond it \citep{berdyugin14}, and begun to reveal its structure. However, with still few measurements, the data within the LHB is sparse and each study monochromatic, meaning we do not know the wavelength dependence of interstellar polarization. Our very best data on wavelength dependence at present comes from \citet{marshall16}, which is based on 2-band measurements of only 4 stars, and gives $\lambda_{max}$ as being in the range 35~nm to 600~nm, with a most probable value of 470~nm.

The LHB is the result of shock waves and stellar winds from ancient supernova explosions sweeping up gas and dust as they expand \citep{frisch17}. \citet{berghofer02} estimate that there have been 20 supernova explosions in this region of space within the last 10--20~Myr. Shock waves can sweep-up and heat the interstellar medium returning thermally volatile components of dust to the gas phase. Consequently the LHB is irregular in shape, inhomogeneous and is made up of a number of features related to these past explosions. One prominent feature is known as the Loop I Superbubble \citep{frisch14}, or simply Loop I. The Loop I Superbubble results from explosions in the ScoCen association $\sim$15~Myr ago. If Loop I is a spherical feature, the Sun sits on or near its rim \citep{frisch90,heiles98,frisch17}. Optical polarization and reddening data show that the eastern parts of Loop I, $l=3-60$, $b>0$, fall within 60--80~pc of the Sun \citep{frisch11,santos11}. The value of $\lambda_{max}$ in the LHB is of particular interest because of this history. The tendency for higher $\lambda_{max}$ in denser regions is thought to be a result of processes which increase average grain size over time, such as coagulation and formation of ice mantles \citep{whittet92}. The LHB is an environment in which these processes are likely in their initial stages or incomplete \citep{voshchinnikov14}, which offers an opportunity to study the properties of grains post supernova shock and in the initial stages of growth.

It is not just the size distribution that is expected to be affected by past supernovae in the LHB, but the composition of the dust as well. In observations of 23 Orionis \citet{welty99} shows that carbonaceous grains are destroyed in shocked low density gas. \citet{slavin08} modelled the Local Interstellar Cloud -- a region within the LHB containing the Sun -- based on short wavelength emission spectra, finding that grains in this region are likely to be olivines with, by contrast, all of the carbon in the gas phase.
 
As well as being of interest for what it can tell us about the ISM, interstellar polarization is an obfuscating addition to any measurement we make of intrinsic polarimetric phenomena. Because interstellar polarization is low within the LHB we can probe weakly polarizing stellar phenomena by observing stars inside it. Some examples of recent work with this \textit{modus operandi} include measuring scattering from the oblate atmosphere of rapidly rotating star Regulus \citep{cotton17b}, an investigation of intrinsic polarization in FGK dwarfs \citep{cotton17a}, and an investigation of magnetic effects in the star $\xi$~Boo~A \citep{cotton18a}. In each case the target object(s) were all very close, within ~25 pc. At distances much further than this the separation of intrinsic and interstellar polarization has been more challenging (cf. \citet{bailey10,cotton16a,marshall16}). One challenge in subtracting interstellar polarization is that its magnitude is not consistent with wavelength. Making many extra control star measurements in additional wavebands is time consuming. General knowledge of the wavelength dependence of interstellar polarization within the LHB is therefore desirable.

In this work we present improved measurements of interstellar polarization in the LHB. In Section~\ref{sec:obs} we describe our instrumentation and observations. In section~\ref{sec:raa} we present multi-wavelength polarization measurements of the nearby, bright stars that comprise our sample, and fit this data with a Serkowski curve and deduce the wavelength of maximum polarization of the Interstellar Medium within the Local Hot Bubble. In section~\ref{sec:discussion} we put the observations in context. Finally, in section~\ref{sec:conclusions} we summarise our findings and present our conclusions.

\section{Observations}
\label{sec:obs}

The observations for this work were carried out with the HIgh Precision Polarimetric Instrument (HIPPI) on the 3.9~m Anglo-Australian Telescope (AAT) at Siding Spring in Australia. The observations were acquired over two runs spanning 2017-06-22 to 2017-07-05 and 2017-08-07 to 2017-08-20.

\subsection{Sample Stars}
\label{sec:stars}

\begin{table*}
	\centering
	\caption{Stellar properties.}
	\label{tab:stars}
	\tabcolsep 3.5 pt
	\begin{tabular}{rrrrrrrrlrrcc} 
		\hline
		Target & \multicolumn{1}{c}{RA} & \multicolumn{1}{c}{Dec} & \multicolumn{1}{c}{l} & \multicolumn{1}{c}{b} & \multicolumn{1}{c}{d$^G$} & \multicolumn{1}{c}{$m_V$} & \multicolumn{1}{c}{$B-V$} & Sp. Type & \multicolumn{1}{c}{$F_{IR}$} & \multicolumn{1}{c}{$v$sin$i$} & \multicolumn{1}{c}{Na~I D$^L$}    &   \multicolumn{1}{c}{Pred. $p_{g^{\prime}}$$^C$}\\
		HD & \multicolumn{1}{c}{(d m s)} & \multicolumn{1}{c}{(d m s)} & \multicolumn{1}{c}{($\degr$)} & \multicolumn{1}{c}{($\degr$)} & \multicolumn{1}{c}{(pc)} & & & & \multicolumn{1}{c}{($\times10^{-6}$)} & \multicolumn{1}{c}{(km/s)} & \multicolumn{1}{c}{(m{\AA})}   &   \multicolumn{1}{c}{(ppm)}\\
		\hline
		4150$^b$       & 00 43 21.2 & $-$57 27 47 & 305.1 & $-$51.6 & 76 & 4.37 &  0.00 & A0~IV    & 1.0$^{CS}$ & 124$^R$ &    $\sim$20    &   113 \\
		17566$^{\0}$   & 02 45 32.6 & $-$67 37 00 & 287.8 & $-$46.0 & 92 & 4.83 &  0.07 & A2~IV/V  &            &  90$^D$ &    $\sim$20$^a$&   139 \\
		125473$^b$     & 14 20 33.4 & $-$37 53 07 & 321.7 & +21.7 & 69 & 4.03 & $-$0.03 & A0~IV    & 1.7$^{Ch}$ & 124$^R$ &    >20, <50    &   101 \\
		138905$^{\0}$  & 15 35 31.6 & $-$14 47 22 & 351.5 & +32.2 & 47 & 3.91 &  1.02 & G8.5~III &            &         &    <20         &   \012 \\
		165760$^{\0}$  & 18 07 18.4 & +08 44 02 & 035.8 & +13.7 & 76 & 4.65 &  0.92 & G8~III   &            &         &    >20, <50    &   113 \\
		206453$^{\0}$  & 21 42 39.5 & $-$18 51 59 & 033.4 & $-$46.0 & 92 & 4.72 &  0.86 & G8~III   &            &         &    <20         &   139 \\
		216735$^{\0}$  & 22 55 13.7 & +08 48 58 & 080.8 & $-$44.3 & 84 & 4.90 &  0.00 & A1~V     &            & 107$^R$ &    >20, <50    &   126 \\
		\hline
	\end{tabular}
	\begin{flushleft}
	References: (G) \citet{gaia18}, (L) \citet{lallement03}, (C) \citet{cotton17a}, (CS) \citet{cotten&song16}, (R) \citet{royer07}, (D) \citet{diaz11}, (Ch) \citet{chen14}. \\
	Notes: \\
	(a) The 20~m{\AA} and 50~m{\AA} contour lines nearly overlap at this point. \\
	(b) HD~4150 and HD~125473 have secondary components, see the text for details.
	\end{flushleft}
\end{table*}

Details of the stars observed in this study are given in table \ref{tab:stars}. Our aim in target selection was to choose stars that allowed us to best sample the wavelength dependence of interstellar polarization within the LHB. This required bright stars (anything much fainter than $V=5$ was likely to be challenging) near the edge of the LHB in order to maximise the difference in polarization between bands whilst minimising uncertainty. Six of the seven targets we observed lie between 65 and 100 pc\footnote{The seventh target, HD~138905, is 47~pc distant; it was originally observed as a control in another study but as its polarization magnitude and spectral type was similar to the other objects observed we added it to this study.} and all seven lie near the edge of the LHB according to the Na~I gas maps of \citet{lallement03}. An estimation of the equivalent Na~I~D line width for each star is given in table \ref{tab:stars}; this comes from the maps presented in \citet{lallement03} where contour lines are drawn for 20~m{\AA} and 50~m{\AA} in 15$\degr$ slices of space. For the distance of our targets we initially estimated the magnitude of polarization in the SDSS g$^{\prime}$ band would be between $\sim$100~ppm and 150~ppm (exact figures are given in the final column of table \ref{tab:stars}) based on the relations given in \citet{cotton16a,cotton17a}, and based on an assumed $\lambda_{max}$ of 470~nm, the difference in polarization between extreme bands would be $\sim$15~ppm to 25~ppm.

With the difference in polarization between bands expected to be so small it was imperative to identify bright targets the least likely to be intrinsically polarized. To that end we avoided active stars,  B-type stars and giants of K-type or later \citep{cotton16a,cotton16b}. This left us with a small number of possible targets, almost exclusively A-type stars and late G-type giants. Either of these stellar types might still appear intrinsically polarized if they host a debris disk \citep{bailey10,cotton16a,cotton17a}, or a close hot-Jupiter exoplanet \citep{bailey18,seager00}. Two of the selected targets have a far-infrared excess consistent with the presence of a debris disk; HD~125473 has a fractional excess of 1.7$\times10^{-6}$ \citep{chen14}, and HD~4150 has a fractional excess of 1.0$\times10^{-6}$ \citep{cotten&song16}, however the polarization that might be expected from such excesses is less than 1~ppm. None of the targets are known to host problematic close planetary companions. 

Close stellar companions can also result in intrinsic polarization. For very short-period binary systems including an early-type star like, for instance, Spica, polarization easily detectable by HIPPI might have gone unidentified in the past \citep{bailey17}. In close companions entrained gas between the two stars creating an asymmetric scattering medium is the accepted polarimetric mechanism; this usually happens as a result of mass transfer between the components \citep{brown78, rudy78} -- this polarization mechanism has no significant wavelength dependence (See \citealp{berdyugin18} for a recent example). The mechanism has only been observed where one of the components is an O- or B-type star, usually early B-type (e.g. \citealp{rudy76, kemp77}) -- but with improved sensitivity it is conceivable examples might be found in later types. Two stars in our survey are known to have close companions: HD~4150 and HD~125473. For HD~4150 its companion, identified using infrared interferometry, is thought to be a K0~V type at 6.8~au separation and a minimum period of 9.8~yr \citep{marion14}. The inclination of the companion's orbit is not known, but the system was previously classified as an astrometric binary \citep{makarov05, frankowski07}, suggesting an orbit that is more face-on. Polarimetry is sensitive to face-on systems. However, binary systems known to exhibit intrinsic polarization have shorter periods, on the order of a week rather than years. Consequently HD~4150 is not expected to display intrinsic polarization on account of its companion. HD~125473 is a spectroscopic binary, the masses of the two stars have been determined through radial velocity measurements to be approximately 3.1 and 1.9~M$_{\sun}$ \citep{mantegazza10}. The eccentricity of the system is quite high at 0.55, but the 38.8~d period \citep{mantegazza10} is probably still too long to produce a significant effect, providing eclipse is avoided\footnote{Intrinsic polarization from an eclipsing binary system was first predicted by \citet{chandrasekhar46}, and has been observed only for Algol \citep{kemp83}.}. We cautiously included HD~125473 on account of no similarly bright better targets being available at the time, but determined to scrutinise the measurements closely for high levels of polarization or deviation from any trends seen.

The A-type stars might also exhibit polarization if they are rotating at close to critical velocity in the same way as Regulus does \citep{cotton17b}. All four of the A-type targets have a published $v$sin$i$, but none is greater than 124~km/s (see table \ref{tab:stars}) -- less than half of the critical velocity for break-up. It is possible that any of these targets might be highly inclined, and rapidly rotating, but there is no way of knowing this in advance. In any case, higher inclinations necessarily result in less observable polarization \citep{cotton17b,sonneborn82}. Thus if polarization due to rapid rotation is present in any of these stars it is likely to be less than that seen in Regulus ($\sim$40~ppm), owing to a cooler temperature and greater inclination. Though where there is a lower gravity that might result in a higher polarization.

\subsection{Instrumentation and Calibration}

\subsubsection{HIPPI}
\label{sec:hippi}

HIPPI is a high precision polarimeter, with a reported sensitivity in fractional polarization of $\sim$4.3~ppm on stars of low polarization, and a precision of better than 0.01~\% on highly polarized stars \citep{bailey15}. It achieves this by the use of a Boulder Nonlinear Systems (BNS) Ferroelectric Liquid Crystal (FLC) modulator operating at a frequency of 500~Hz to eliminate the effects of variability in the atmosphere. Second stage chopping, to reduce systematic effects, is accomplished by rotating the entire back half of the instrument after the filter wheel, with a typical frequency of once per 20~s. And each target is measured at four different position angles (PAs): 0, 45, 90 and 135 degrees, where the redundant angles facilitate the easy removal of instrumental effects.

For the majority of observations, a sky measurement, lasting 40~s, was acquired at each of the four telescope position angles an object was observed at, and subtracted from the measurement. In good sky conditions, for some red band observations a dark measurement was sufficient for calibration purposes. These subtractions were carried out as the first part of the data reduction routine, that determines polarization via a Mueller Matrix method. Full details are provided by \citet{bailey15}.

In this work data were acquired in four filters. Observations in SDSS $g^{\prime}$ and a 425~nm short pass (425SP) filter were made using blue sensitive Hamamatsu H10720-210 Ultra bialkali photocathode (B) photomultiplier tubes (PMTs). Observations made in SDSS $r^{\prime}$ and a 650~nm long pass (650LP) filter were made using red sensitive Hamamatsu H10720-20 infrared extended multialkali photocathodes (R); which have some sensitivity out to $\sim$900~nm. A bandpass model, as described in \citet{bailey15} is used to determine the effective wavelengths and modulation efficiency correction (discussed below) to the raw data for each filter/PMT combination. The modulation efficiency and effective wavelength are explicitly calculated for spectral types O3, B0, A0, F0, G0, K0, M0 and M5 using Kurucz model spectra for dwarfs of those types; the results are interpolated for intermediate types. In this work our targets are relatively close and no reddening has been applied in the bandpass model. 

HIPPI has now been superseded by HIPPI-2 (paper forthcoming), for the new instrument we made improvements to the reduction software and bandpass model which we have also implemented here for HIPPI. An improvement in the data reduction software was made by increasing the numerical precision in the calibration routines. There are two main improvements in the bandpass model. The first is that we have improved our characterisation of the optical components. We now include the spectral response (transmission or reflection) of every optical component in the light path, including the telescope mirrors. For the components in the instrument we have supplemented the supplied manufacturer data with lab based measurements using a Cary 1E UV-Visible spectrometer. This means there is no longer an error resulting from manufacturing tolerances in the transmission of some lenses and filters. This is important because the full passband of HIPPI is very broad, and at the extremes outside the designed wavelength range of some components. The second improvement is that we now consider airmass (to 0.1~atm precision) when calculating atmospheric transmission.

The improvements in the bandpass model have allowed us to use multi-band data collected on high polarization standards to update the modulator calibration we reported in \citet{bailey15}. To achieve that we carry out $\chi^2$ fitting of the modulator's performance parameters using a full bandpass model (not just the effective wavelengths). The modulation efficiency is given by the equation: \begin{equation}e(\lambda) = \frac{e_{max}}{2}\left ( 1-\cos2\pi\frac{\Delta}{\lambda} \right ),\end{equation} where $e_{max}$ is the maximum efficiency, $\lambda$ is the wavelength. The path difference between the two optical axes of the retarder, $\Delta$, is given by: \begin{equation}\Delta = \frac{\lambda_0}{2}+Cd\left ( \frac{1}{\lambda^2}- \frac{1}{\lambda_0^2} \right ),\end{equation} where $\lambda_0$ is the wavelength corresponding to peak efficiency, and the two terms $Cd$ can be treated as one constant (C is a parameter describing the dispersion in birefringence of the FLC material and d is the thickness of the FLC layer). We've found the BNS FLC's performance characteristics can drift slowly over time, such that $\lambda_0$ is $\sim$10~nm redder for the data taken in 2016/2017 than for 2014/2015. This difference is inconsequential for the majority of our observations, but in the 425SP band can make a difference. From 15 observations of our regular high polarization standards and $\zeta$~Oph in 6 different bands for 2016/2017 we calculate $\lambda_0$ to be 506.6~$\pm$~2.9~nm, and $Cd$ to be (1.758~$\pm$~0.116)~$\times10^7$~nm$^3$.

The output from our standard bandpass models have been previously reported, and as this is little different for g$^{\prime}$ and r$^{\prime}$ filters, we do not repeat that information here (see the supplementary information in \citealp{cotton17b} for example). We also give the exact calculated modulation efficiency for each observation in sections \ref{sec:patp} and \ref{sec:results}. The 650LP filter is used here with HIPPI for the first time, and we give the details of the bandpass model for it, and for 425SP with two different airmasses (1~atm and 2~atm) below in table \ref{tab:bpeff}. The band modulation efficiency being: \begin{equation}\mathrm{Eff.}=\frac{\int e(\lambda)S(\lambda).d\lambda}{\int S(\lambda).d\lambda},\end{equation} where $S(\lambda)$ is the relative contribution to the output detector signal as a function of wavelength. All observations are corrected by dividing by $\mathrm{Eff.}$ Given for reference is the effective wavelength of the band: \begin{equation}\lambda_{\rm eff}=\frac{\int S(\lambda).d\lambda}{\int S(\lambda).d\lambda}.\end{equation}

\begin{table}
    \centering
    \tabcolsep 8 pt
	\centering
	\caption{650LP and 425SP effective wavelength and modulation efficiency.}
	\label{tab:bpeff}
	\begin{tabular}{lrrrrrr} 
		\hline
		Spectral & \multicolumn{2}{c}{650LP AM1} & \multicolumn{2}{c}{425SP AM1} & \multicolumn{2}{c}{425SP AM2} \\
		Type     & $\lambda_{\rm eff}$ & Eff. & $\lambda_{\rm eff}$ & Eff. & $\lambda_{\rm eff}$ & Eff. \\
		     & (nm)                & (\%) & (nm)                & (\%) & (nm)                & (\%) \\
		\hline
        O3  &   714.8   &   66.3    &   393.7   &   45.6    &   395.1   &   46.8    \\
        B0  &   715.2   &   66.2    &   394.2   &   46.0    &   395.5   &   47.2    \\
        A0  &   718.3   &   65.8    &   400.3   &   51.8    &   401.4   &   52.7    \\
        F0  &   720.7   &   65.4    &   400.6   &   51.5    &   402.0   &   52.4    \\
        G0  &   722.7   &   65.1    &   401.4   &   51.2    &   403.3   &   52.3    \\
        K0  &   724.0   &   64.9    &   404.8   &   52.7    &   407.3   &   53.8    \\
        M0  &   737.5   &   62.9    &   414.2   &   54.9    &   418.7   &   55.9    \\
        M5  &   742.0   &   62.2    &   414.5   &   54.9    &   419.0   &   55.9    \\
		\hline
	\end{tabular}
\end{table}

\subsubsection{Position Angle and Telescope Polarization Calibration}
\label{sec:patp}

\begin{table}
	\centering
	\caption{Telescope polarization calibration.}
	\label{tab:tp}
	\tabcolsep 3 pt
	\begin{tabular}{rrrrr} 
		\hline
		HD      & \multicolumn{1}{c}{UTC} & \multicolumn{1}{c}{Eff.} & \multicolumn{1}{c}{$q$} & \multicolumn{1}{c}{$u$}        \\
		        &                     & & \multicolumn{1}{c}{(ppm)}      & \multicolumn{1}{c}{(ppm)}      \\
		\hline
		\textbf{425SP} \\
        2151    & 2017-06-22 19:45 & 0.522 &	$-$47.6 $\pm$	11.8 &	21.0 $\pm$ 11.0 \\
        2151    & 2017-06-25 19:13 & 0.523 &	$-$10.8 $\pm$	12.7 &	16.1 $\pm$ 12.7 \\
        48915   & 2017-08-11 19:21 & 0.501 &	 37.0 $\pm$	10.1 &	57.7 $\pm$ 10.8 \\
        48915   & 2017-08-12 19:41 & 0.498 &	$-$13.4 $\pm$ \01.7 &  27.1 $\pm$ \01.8 \\
        48915	& 2017-08-19 19:17 & 0.498 &	$-$10.2 $\pm$ \08.1 & $-$33.6 $\pm$	\06.9 \\
        102647  & 2017-06-22 09:04 & 0.499 &	 14.5 $\pm$	\07.0 &   8.1 $\pm$ \06.8 \\
        102647  & 2017-06-30 08:59 & 0.500 &	$-$22.8 $\pm$ \07.3 & $-$49.2 $\pm$	\08.1 \\
        102870  & 2017-06-23 08:58 & 0.520 &	 19.9 $\pm$ 15.6 &  23.5 $\pm$ 15.6 \\
        102870	& 2017-07-01 09:10 & 0.521 &	$-$32.9 $\pm$ 16.3 &   2.2 $\pm$ 16.2 \\
        &&\textbf{ TP:}	              &  \textbf{$-$7.4 $\pm$\03.7} &	 \textbf{8.1 $\pm$\03.6}\\
        \hline
        \textbf{SDSS g$^{\prime}$} \\
		2151    & 2017-06-25 19:36 & 0.888 &	$-$21.3 $\pm$	\04.2 &   6.8 $\pm$ \04.1 \\
        2151    & 2017-08-10 19:06 & 0.888 &	$-$16.9 $\pm$	\04.2 &   4.4 $\pm$ \04.6 \\
        48915   & 2017-08-11 19:41 & 0.872 &	$-$21.4 $\pm$	\04.8 & $-$10.3 $\pm$ \05.0 \\
        48915   & 2017-08-19 19:01 & 0.872 &	  2.6 $\pm$	\02.7 &  $-$5.7 $\pm$ \02.6 \\
        102647  & 2017-06-22 09:04 & 0.873 &	 $-$3.1 $\pm$	\02.4 &   0.7 $\pm$ \02.6 \\
        102647  & 2017-06-30 08:27 & 0.873 &	 $-$4.7 $\pm$	\02.5 & $-$19.9 $\pm$ \02.5 \\
        102870  & 2017-06-23 08:58 & 0.886 &	$-$10.9 $\pm$	\05.2 &  15.5 $\pm$ \04.9 \\
        102870  & 2017-06-25 08:23 & 0.886 &	 $-$3.1 $\pm$	\05.9 & $-$10.5 $\pm$ \05.4 \\
        &&\textbf{ TP:}                &  \textbf{$-$9.9 $\pm$\01.5} &	\textbf{$-$2.4 $\pm$\01.5}\\
        \hline
        \textbf{SDSS r$^{\prime}$}	\\						
        2151	& 2017-07-04 14:25 & 0.816 &	$-$18.8 $\pm$	\04.5 &	$-$9.0 $\pm$ \04.7 \\
        2151	& 2017-08-09 12:16 & 0.816 &	$-$21.1 $\pm$ \04.3 &   3.5 $\pm$ \04.3 \\
        2151	& 2017-08-09 19:04 & 0.817 &	$-$24.1 $\pm$ \04.1 &  $-$8.4 $\pm$ \04.0 \\
        48915	& 2017-08-07 20:00 & 0.824 &    $-$18.6 $\pm$ \02.3 &	$-$7.8 $\pm$ \02.3 \\
        48915	& 2017-08-08 19:55 & 0.824 &    $-$11.9 $\pm$ \01.9 &	$-$8.1 $\pm$ \01.5 \\
        48915	& 2017-08-09 19:30 & 0.824 &    $-$11.7 $\pm$ \01.2 &	$-$7.8 $\pm$ \01.4 \\
        102647	& 2017-07-04 08:07 & 0.824 &      2.8 $\pm$ \03.5 &  $-$4.1 $\pm$ \03.6 \\
        102647	& 2017-07-04 08:34 & 0.823 &     $-$4.5 $\pm$ \03.6 &	$-$4.9 $\pm$ \03.6 \\
        102870	& 2017-07-04 09:10 & 0.817 &     $-$0.8 $\pm$ \06.0 & $-$13.9 $\pm$ \05.9 \\
        102870	& 2017-07-04 09:44 & 0.817 &      7.3 $\pm$ \06.1 & $-$12.7 $\pm$ \05.8 \\
        &&\textbf{ TP:}	              & \textbf{$-$10.1 $\pm$\01.3} &	\textbf{$-$7.3 $\pm$\01.3}\\
		\hline
		\textbf{650LP} \\
        2151	& 2017-07-04 14:25 & 0.646 &	$-$14.6 $\pm$ \06.7 &	 2.0 $\pm$ \07.0 \\
        2151	& 2017-08-08 19:25 & 0.646 &	  3.1 $\pm$ \07.2 &  $-$9.3 $\pm$ \06.9 \\
        2151	& 2017-08-09 12:41 & 0.646 &	$-$14.5 $\pm$ \07.1 &	 7.1 $\pm$ \07.2 \\
        48915	& 2017-08-07 19:43 & 0.658 &	$-$11.1 $\pm$ \04.6 &  $-$9.5 $\pm$ \05.6 \\
        48915	& 2017-08-08 19:55 & 0.658 &	 $-$8.3 $\pm$ \02.7 & $-$10.4 $\pm$ \02.4 \\
        48915	& 2017-08-09 19:47 & 0.658 &	 $-$4.9 $\pm$ \01.9 &	$-$7.6 $\pm$ \02.1 \\
        102870	& 2017-07-04 09:10 & 0.647 &	 $-$8.1 $\pm$ \09.9 & $-$10.3 $\pm$ \09.8 \\
        &&\textbf{ TP:}	              &  \textbf{$-$8.4 $\pm$\02.4} &	\textbf{$-$5.4 $\pm$\02.4}\\
		\hline
	\end{tabular}
\end{table}

Angular calibration was carried out with reference to a set of high polarization standards (observations in g$^{\prime}$ with known polarization angles. For the 2017-06-22 to 2017-07-05 run the set was: ($\times$2) HD~147084 (32.0$\degr$), HD~154445 (90.1$\degr$), and HD~160529 (20.4$\degr$); with the standard deviation of those measurements being 1.1$\degr$. For the 2017-08-07 to 2017-08-20 run the set was: HD~147084, HD~154445 and HD~187929 (93.8$\degr$); with the standard deviation of those measurements being 0.5$\degr$. The standards have angles known to a precision of $\sim1\degr$ -- which is the main source of uncertainty in these measurements. The observations were $PA$ (position angle) corrected before any other corrections were applied. During the observations, zero point calibration (telescope polarization; hereafter abbreviated TP) was carried out in each filter/PMT combination by reference to the average of a set of observed stars with measured low polarizations; this is shown in Table \ref{tab:tp} in terms of Stokes parameters $q=Q/I$ and $u=U/I$. The scatter in the observations of Sirius (HD~48915) in 425SP and g$^{\prime}$ is larger than usual. We ascribe this to particularly poor seeing (around 5~$\arcsec$ or worse) combined with a low observing angle at the time of those observations. It is worth noting that essentially the same standards have been observed in each band. We have a high degree of confidence in the TP determinations since the TP is particularly low in every band and that the differences between bands are small. It is worth noting that the 425SP TP in $u$ is not as consistent with the other bands, however it is normal for there to be variations in the TP with wavelength.

\section{Results and Analysis}
\label{sec:raa}

\subsection{Results}
\label{sec:results}

\begin{table*}
	\centering
	\caption{Polarimetry of sample stars.}
	\label{tab:res}
	\begin{tabular}{lcccrrrrrrr} 
		\hline
		UTC                  & Obs. & Sky & Fil. &  \multicolumn{1}{c}{$\lambda_{\rm eff}$} & \multicolumn{1}{c}{Eff.} & \multicolumn{1}{c}{Exp.}  & \multicolumn{1}{c}{$q$}     & \multicolumn{1}{c}{$u$}      & \multicolumn{1}{c}{$\hat{p}$}        & \multicolumn{1}{c}{$PA$} \\
		                     &      &     &      & \multicolumn{1}{c}{(nm)}             && \multicolumn{1}{c}{(s)}   & \multicolumn{1}{c}{(ppm)} & \multicolumn{1}{c}{(ppm)}  & \multicolumn{1}{c}{(ppm)}    & \multicolumn{1}{c}{($\degr$)} \\   
		\hline
		\textbf{HD~4150} \\
		2017-07-05 15:20:25	& 1 & S & 425SP        & 399.8 & 0.496 &    2560 & $-$110.3 $\pm$ \09.2 &   13.7 $\pm$ \09.2 & 110.7 $\pm$ \09.2 &  86.5 $\pm$ 2.4 \\
        2017-07-01 18:48:18	& 1	& S & g$^{\prime}$ & 464.6 & 0.867 &     720 &  $-$80.8 $\pm$ \07.6 &    4.8 $\pm$ \07.8 &  80.5 $\pm$ \07.7 &  88.3 $\pm$ 2.8 \\
        2017-07-04 15:00:43	& 1	& D & r$^{\prime}$ & 620.5 & 0.824 &    1280 &  $-$68.6 $\pm$ \07.0 &    6.1 $\pm$ \07.0 &  68.4 $\pm$ \07.0 &  87.5 $\pm$ 2.9 \\
        2017-08-08 18:36:54	& 1	& S & 650LP        & 717.4 & 0.659 &    2560 &	 $-$85.9 $\pm$ \08.8 &   23.7 $\pm$ \08.9 &  88.7 $\pm$ \08.8 &  82.3 $\pm$ 2.9 \\
		\hline
        \textbf{HD~17566} \\
        2017-08-12 09:57:44	& 2$^a$ & S & 425SP	   & 399.7 & 0.496 &    5120 &  $-$43.2 $\pm$ \09.1 &  $-$70.5 $\pm$ \09.1 &  82.1 $\pm$ \09.1 & 119.2 $\pm$ 3.2 \\
        2017-08-11 15:37:08	& 1	& S & g$^{\prime}$ & 466.2 & 0.872 &    1600 &  $-$38.9 $\pm$ \05.7 &  $-$43.3 $\pm$ \06.5 &  57.9 $\pm$ \06.1 & 114.0 $\pm$ 3.0 \\
        2017-08-09 17:17:49	& 1	& S & r$^{\prime}$ & 620.7 & 0.824 &    2560 &  $-$45.0 $\pm$ \05.8 &  $-$44.2 $\pm$ \05.7 &  62.8 $\pm$ \05.7 & 112.3 $\pm$ 2.6 \\
        2017-08-09 17:17:34	& 2$^b$	& S & 650LP	   & 717.9 & 0.657 &    5120 &  $-$43.5 $\pm$ \07.4 &  $-$44.4 $\pm$ \07.5 &  61.8 $\pm$ \07.4 & 112.8 $\pm$ 3.4 \\
		\hline
		\textbf{HD~125473}* \\
		2017-07-05 09:21:12	& 1 & S & 425SP	       & 399.0 & 0.489 &    2560 &	$-$126.5 $\pm$ \08.4 &  $-$90.4 $\pm$ \07.9 & 155.3	$\pm$ \08.2 & 107.8 $\pm$ 1.5 \\
        2017-06-29 08:38:48	& 1 & S & g$^{\prime}$ & 464.3 & 0.867 &     800 &	 $-$79.1 $\pm$ \07.0 & $-$111.4 $\pm$ \07.5 & 136.4	$\pm$ \07.3 & 117.3 $\pm$ 1.5 \\
        2017-07-04 10:17:00	& 1 & S & r$^{\prime}$ & 620.1 & 0.825 &    1800 &	 $-$87.5 $\pm$ \04.8 &  $-$86.2 $\pm$ \05.2 & 122.7	$\pm$ \05.0 & 112.3 $\pm$ 1.2 \\
        2017-08-08 08:59:33	& 1 & S & 650LP	       & 717.4 & 0.659 &    2240 &	 $-$75.7 $\pm$ \08.2 &  $-$80.3 $\pm$ \08.1 & 110.1	$\pm$ \08.2 & 113.3 $\pm$ 2.1 \\
		\hline
		\textbf{HD~138905} \\
		2017-07-04 01:28:43	& 2$^c$	& S & 425SP    & 405.5 & 0.532 &    4480 &	 $-$18.2 $\pm$ 11.6  &  156.3 $\pm$ 11.2 & 157.0 $\pm$ 11.4 &	   48.3 $\pm$ 2.1 \\
        2017-06-23 11:59:01	& 1	& S & g$^{\prime}$ & 474.1 & 0.894 &     640 &	 $-$68.3 $\pm$ \06.5 &  131.4 $\pm$ \06.4 & 147.9	$\pm$ \06.4 &  58.7 $\pm$ 1.3 \\
        2017-07-04 11:10:43	& 1	& S & r$^{\prime}$ & 626.4 & 0.814 &    1920 &	 $-$49.0 $\pm$ \03.9 &  127.0 $\pm$ \03.8 & 136.0	$\pm$ \03.8 &  55.6 $\pm$ 0.8 \\
        2017-08-08 09:57:58	& 1	& S & 650LP        & 725.1 & 0.641 &    2560 &	 $-$36.3 $\pm$ \05.4 &  103.8 $\pm$ \05.5 & 109.8	$\pm$ \05.4 &  54.6 $\pm$ 1.4 \\
        \hline
        \textbf{HD~165760} \\
        2017-07-01 13:39:54	& 1 & S & 425SP	       & 405.6 & 0.533 &    2640 &	 433.1 $\pm$ 17.1  &  603.8 $\pm$ 16.7  & 742.9 $\pm$ 16.9  &  27.2 $\pm$ 0.7 \\
        2017-07-01 14:23:54	& 1 & S & g$^{\prime}$ & 474.5 & 0.895 &     640 &	 511.7 $\pm$ \07.9 &  673.9 $\pm$ \08.1 & 846.1 $\pm$ \08.0 &  26.4 $\pm$ 0.3 \\
        2017-07-04 11:50:25	& 1 & D & r$^{\prime}$ & 626.5 & 0.813 &     800 &	 552.8 $\pm$ \08.2 &  628.2 $\pm$ \08.3 & 836.8 $\pm$ \08.2 &  24.3 $\pm$ 0.3 \\
        2017-08-08 10:55:44	& 1 & S & 650LP	       & 724.9 & 0.641 &    2080 &	 518.9 $\pm$ \08.5 &  613.3 $\pm$ \08.3 & 803.4 $\pm$ \08.4 &  24.9 $\pm$ 0.3 \\
        \hline
        \textbf{HD~206453} \\
        2017-07-05 14:21:04	& 1	& S & 425SP	       & 405.3 & 0.532 &    2560 &  $-$41.0 $\pm$ 17.1  &  $-$28.9 $\pm$ 17.6  &  47.1 $\pm$ 17.3 &	107.6 $\pm$10.7 \\
        2017-07-04 19:40:18	& 1	& S & g$^{\prime}$ & 474.5 & 0.895 &     800 &  $-$87.7 $\pm$ \08.0 &   33.0 $\pm$ \07.7	&  93.4 $\pm$ \07.8 &	 79.7 $\pm$ 2.4 \\
        2017-07-04 13:44:02	& 1	& D & r$^{\prime}$ & 626.5 & 0.813 &    1280 &  $-$68.1 $\pm$ \06.7 &   31.8 $\pm$ \06.7	&  74.9 $\pm$ \06.7 &	 77.5 $\pm$ 2.6 \\
        2017-08-09 11:26:58	& 1	& S & 650LP	       & 724.9 & 0.641 &    2560 &  $-$67.3 $\pm$ \07.9 &   22.7 $\pm$ \07.9	&  70.6	$\pm$ \07.9 &	 80.7 $\pm$ 3.2 \\
        \hline
        \textbf{HD~216735} \\
        2017-08-12 17:04:40	& 1	& S & 425SP	       & 399.7 & 0.495 &    2560 &	$-$689.3 $\pm$ 12.4  & $-$201.5 $\pm$ 11.5  & 718.0 $\pm$ 12.0  &  98.1 $\pm$ 0.5 \\
        2017-08-12 16:26:17	& 1	& S & g$^{\prime}$ & 465.3 & 0.870 &     800 &	$-$739.6 $\pm$ \07.5 & $-$198.9 $\pm$ \07.1 & 765.9	$\pm$ \07.3 &  97.5 $\pm$ 0.3 \\
        2017-08-09 15:36:23	& 1	& S & r$^{\prime}$ & 620.5 & 0.825 &     960 &	$-$745.1 $\pm$ \09.2 & $-$168.6 $\pm$ \09.5 & 763.9	$\pm$ \09.3 &  96.4 $\pm$ 0.4 \\
        2017-08-09 14:26:16	& 1	& S & 650LP	       & 717.6 & 0.658 &    1920 &	$-$723.1 $\pm$ 13.2  & $-$190.8 $\pm$ 12.8  & 747.7	$\pm$ 13.0  &  97.4 $\pm$ 0.5 \\
        \hline
	\end{tabular}
	    \begin{flushleft}
	    Notes: Where multiple observations have been made, data given above are error weighted averages in $q$, $u$ and UT; $p$ and $PA$ have been calculated from the resulting $q$ and $u$. Individual observations follow as UT, Eff., Exp., q, u: \\
	    (a) 2017-08-11 18:31:35, 0.496, 2560, \0$-$3.9 $\pm$ 15.0, $-$73.0 $\pm$ 14.3; 2017-08-12 18:07:05, 0.496, 2560, $-$62.4 $\pm$ 10.8, $-$69.0 $\pm$ 11.0. \\
	    (b) 2017-08-09 16:19:41, 0.657, 2560,  $-$48.7 $\pm$ 10.1, $-$28.9 $\pm$ 10.3; 2017-08-09 18:16:01, 0.657, 2560, $-$38.2 $\pm$ 10.2, $-$60.1 $\pm$ 10.3. \\
	    (c) 2017-06-30 13:43:34, 0.533, 1920,  $-$36.9 $\pm$ 21.3, 121.7 $\pm$ 19.3; 2017-07-05 12:21:29, 0.531, 2560, $-$11.0 $\pm$ 13.5, 172.1 $\pm$ 13.3. \\
	    (*) Observations of HD~125473 do not correspond to the eclipse phase noted in the literature \citep{mantegazza10}.\\
	    \end{flushleft}
\end{table*}

Table~\ref{tab:res} shows the $PA$, TP and efficiency corrected observations of each science target in each filter band. Errors in the determined TP have been incorporated into the errors in $q$ and $u$; the given polarization magnitude has been debiased according to $\hat{p}=\sqrt{p^2-\sigma_p^2}$, where $\sigma_p$ is the error in $p$.

\subsection{Analysis}
\label{sec:analysis}

The primary method of analysis employed in this work is $\chi^2$ fitting of the multi-band measurements to Serkowski curves. For this we use the \textsc{Python} language's `scipy' package \citep{jones01}, specifically the `curve$\_$fit' function. In the first instance each target has a 3- and 2-parameter fit applied to the four-band observations. Rather than simply fitting the measured polarization at the effective wavelength to the Serkowski curve, we use our bandpass model to calculate the polarization magnitude that would be measured in each band if only interstellar polarization was present, and fit our data to that. We use Kurucz stellar spectra for types O3, B0, A0, F0, G0, K0, M0 and M5 in the bandpass model. For spectral types falling between the models we make a calculation for the immediately warmer and cooler types and linearly interpolate the results according to subtypes. The bandpass model also takes into account the airmass at the time of observation. For the targets with multiple observations in a bandpass the airmasses were very similar, so we used the polarization-error weighted-average airmass.  

\begin{figure*}
	\includegraphics[width=0.95\textwidth, trim={0 1 0 1}]{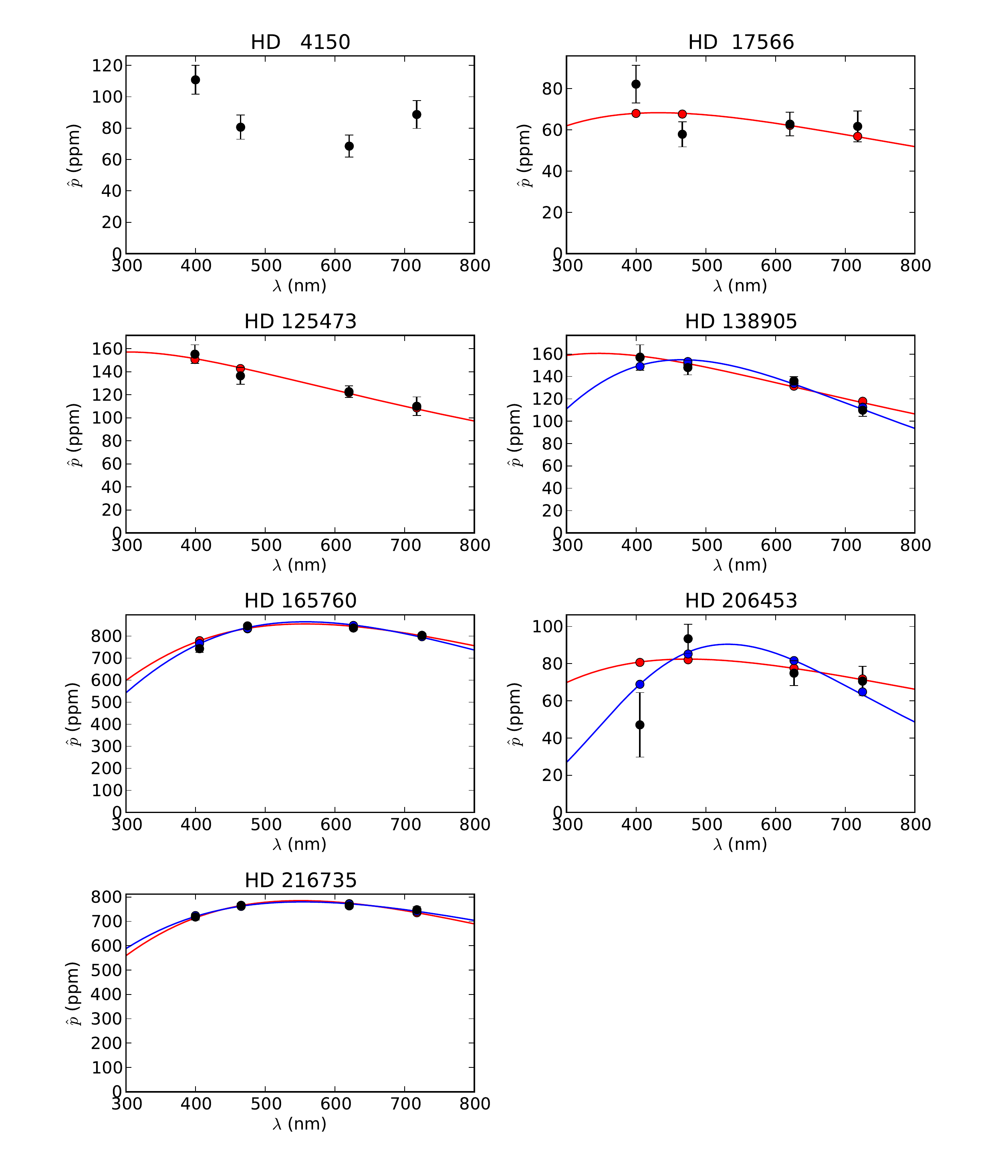}
    \caption{3-parameter (blue line) and 2-parameter (red line) Serkowski fits to the observation of each target. The black points represent the data (error bars are shown in every case), blue and red points represent the bandpass determination corresponding to the data points for the fit parameters. No realistic fit to the observations of HD~4150 was possible, only 2-parameter fits for HD~17566 and HD~125473 were possible. For the two objects exhibiting greater polarization -- HD~165760 and HD~216735 -- the 3-parameter fits are very similar to the 2-parameter fits.}
    \label{fig:3p2p_Serk}
\end{figure*}

To determine $\lambda_{max}$ for each target we then fit 3- and 2-parameter Serkowski functions. Where the 3-parameter fits are for the Serkowski function given by equation \ref{eq:Serk}, and the 2-parameter fits have $K$ determined by $\lambda_{max}$ according to equation \ref{eq:Whit}. The fits are shown in figure \ref{fig:3p2p_Serk}, and the fitted parameters tabulated in table \ref{tab:3p2p_Serk}. Only 2-parameter fits are shown/given for HD~17566 and HD~125473, and no fits are given for HD~4105 at all, as fitting a Serkowski curve did not result in realistic parameters otherwise. 

A $PA$ has also been calculated for each target using $\chi^2$ minimisation, assuming it is identical in each band; this is given in table \ref{tab:3p2p_Serk}. In this case an interstellar polarization model is rotated to reduce the difference between the $q$ and $u$ data and the model (the same result can be achieved by rotating to minimise $u$). 

Two of the targets, HD~165760 and HD~216735, are far more polarized than the others. For HD~165760 and HD~216735 the data is sufficiently constrained that 2-parameter and 3-parameter fits are very similar (they also agree within error); this is owing to the higher polarizations recorded. For the other objects, proportionally the errors are larger, and there is less agreement. In this instance the 3-parameter fits are ill-constrained, and we haven't investigated them further. Indeed, the 3-parameter fits have a worse reduced-$\chi^2$ statistic than the 2-parameter fits in every case. However, excepting HD~4150, the lower polarization objects all display a lower $\lambda_{max}$ than HD~165760 and HD~216735. A reasonable deduction to make is that HD~165760 and HD~216735 are characteristic of dust in or beyond the wall of the LHB, whilst the other targets represent dust within the LHB. 

\begin{table*}
	\centering
	\caption{Single target 2- and 3-parameter Serkowski fits.}
	\label{tab:3p2p_Serk}
	\begin{tabular}{rcccrrccccc} 
		\hline
		        &   \multicolumn{4}{c}{3-parameter}                                          &   \multicolumn{4}{c}{2-parameter}                                         &   &   \\
		HD      &   $p_{max}$       &   $\lambda_{max}$ &   $K$             &   $\chi^2_r$  &   $p_{max}$       &   $\lambda_{max}$   &    $K$*   &   $\chi^2_r$    &   $PA$    \\
		        &   (ppm)           &   (nm)            &                   &               &   (ppm)           &   (nm)                &           &               & ($\degr$)     \\
		\hline
		4105    &                   &                   &                   &               &                   &                     &           &           &   \086.4 $\pm$ 1.8   \\ 
		17566   &                   &                   &                   &               &   \068.3 $\pm$ \09.1 &  432.5 $\pm$ 170.6   &   0.73    &   2.72    &  114.3 $\pm$ 1.5   \\
		125473  &                   &                   &                   &               &   157.1 $\pm$ 12.9  &   305.1 $\pm$ \061.7  &   0.52    &   0.58    &  112.6 $\pm$ 1.3   \\
		138905  &   155.0 $\pm$ \07.2 &   465.7 $\pm$ 52.5  &   1.72 $\pm$ 1.09   &   1.78  &   160.7 $\pm$ 18.5  &   345.5 $\pm$ \089.6  &   0.58    &   1.92    & \055.6 $\pm$ 1.0   \\
		165760  &   865.0 $\pm$ 21.6  &   556.7 $\pm$ 17.3  &   1.22 $\pm$ 0.54   &   6.08  &   855.3 $\pm$ \09.4 &   556.6 $\pm$ \017.9  &   0.93    &   3.91    & \025.4 $\pm$ 0.4   \\
		206453  &   \090.4 $\pm$ 17.1 &   531.4 $\pm$ 58.1  &   3.71 $\pm$ 4.89   &   4.42  &   \082.4 $\pm$ 11.2 &   473.5 $\pm$ 186.4   &   0.80    &   3.00    & \079.8 $\pm$ 2.2   \\
		216735  &   780.1 $\pm$ 10.5  &   553.4 $\pm$ 15.9  &   0.75 $\pm$ 0.28   &   1.38  &   785.2 $\pm$ \05.2 &   550.1 $\pm$ \0\09.8 &   0.92    &   0.94    & \097.3 $\pm$ 0.2   \\
	\hline
	\end{tabular}
	\begin{flushleft}
	Notes: (*) Calculated using equation \ref{eq:Whit}. \\
	\end{flushleft}	
\end{table*}

\begin{figure}
	\includegraphics[width=\columnwidth, trim={0.5cm 0.5cm 0.5cm 0.5cm}]{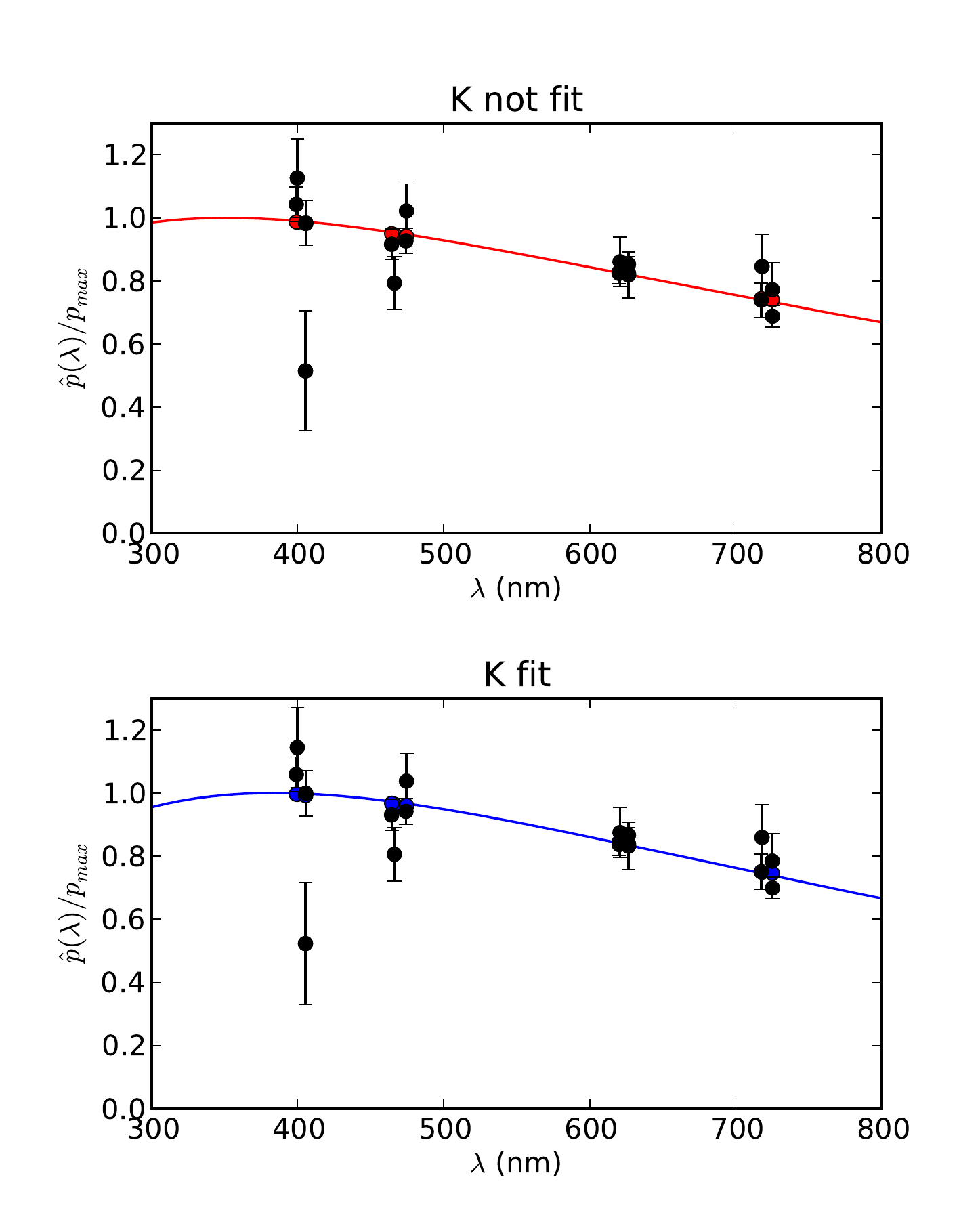}
    \caption{The results of Serkowski fits to four targets (HD~17566, HD~125473, HD~138905 and HD~206453). In both panels $p_{max}$ is fit for each target, along with $\lambda_{max}$, in the lower panel $K$ is also fit. The data points in black are shown as fractional polarization, that is $p$ divided by the fit $p_{max}$ for the target. The fit Serkowski curves are shown in red (without $K$ -- analogous to the 2-parameter fits in figure \ref{fig:3p2p_Serk}) and blue (with $K$ -- analogous to the 3-parameter fits in figure \ref{fig:3p2p_Serk}), and the corresponding coloured points correspond to the fits for the bands corresponding to the data points.}
    \label{fig:comb_Serk}
\end{figure}

HD~165760 and HD~216735 seem to sample a distinctly different dust population to the other stars. Presuming that HD~17566, HD~125473, HD~138905 and HD~206453 sample basically the same dust population we simultaneously fit $p_{max}$ for each of them along with $\lambda_{max}$ and with (and without) $K$. The results are shown in table \ref{tab:comb_Serk}, and in figure \ref{fig:comb_Serk} where the fractional polarization for each data point/target is determined by dividing by the corresponding $p_{max}$ value in the table. We exclude HD~4150 from this analysis since we could not make a fit to it earlier (figure \ref{fig:3p2p_Serk}), and including it here along with the other targets only increased the uncertainties and produced unrealistic results; thus there is likely an intrinsic component to its polarization. We decided to include HD~125473 in spite of its binarity because although the trend with wavelength is a little noisier than the others. The low levels of polarization strongly suggest there is no entrained gas, and the wavelength trend is fairly clearly similar to the other stars.

\begin{table*}
	\centering
	\caption{Multi-target Serkowski fits.}
	\label{tab:comb_Serk}
	\begin{tabular}{cccccclc}
	    \hline
	        Fit $K$ &   \multicolumn{4}{c}{$p_{max}$}                                                      &   $\lambda_{max}$ &   \0\0\0$K$       &   $\chi^2_r$  \\
	                &   \multicolumn{4}{c}{(ppm)}                                                           & (nm)             &                    &               \\
	                &   HD~17566            &   HD~125473           &   HD~138905       &   HD~206453       \\
	                \hline
	        Yes     &    71.8 $\pm$ 6.9       &   146.6 $\pm$ 10.6      &   157.1 $\pm$ 12.0  &   90.0 $\pm$ 9.0    &   383.8 $\pm$ 129.4  &  0.75 $\pm$ 0.73   &   1.82        \\
	        No      &    73.0 $\pm$ 6.4       &   148.9 $\pm$ \09.6     &   159.5 $\pm$ 11.2  &   91.4 $\pm$ 8.5    &   351.1 $\pm$ \053.0  &  0.59*           &   1.67        \\
	\hline
	\end{tabular}
	\begin{flushleft}
	Notes: (*) Calculated using equation \ref{eq:Whit}. \\
	\end{flushleft}	
\end{table*}

\section{Discussion}
\label{sec:discussion}

\subsection{Stars in relation to features in the LHB}
\label{sec:wall}

It is immediately clear from table \ref{tab:res} that two of the targets, HD~165760 and HD~216735, are more polarized than the others -- and more polarized than was initially expected (table \ref{tab:stars}). The $PA$ of both stars is consistent across all four bands, and the data is well fit by a Serkowski curve, confirming the nature of the polarization detected as interstellar. Furthermore $\lambda_{max}$ for both stars is similar to that typical of the ISM beyond the LHB, a fact we explore further in section \ref{sec:colour}. The obvious conclusion to draw is that they represent stars in/beyond the wall of the LHB, whilst the other stars being more weakly polarized are probably within the LHB. 

Using a number of catalogues, \citet{gontcharov17} recently plotted $p$ against distance for 5000 stars out to 650~pc. His plot shows a bifurcation of $p/d$ trends for stars within $\sim$120~pc. Much of \citet{gontcharov17}'s data comes from the \citet{heiles00} catalogue, which itself is an agglomeration of many much older catalogues containing measurements made with less precise instrumentation\footnote{The heritage of some of this data can be traced all the way back to the 1960s or earlier when the first stellar polarimetric measurements were made with photoelectric detectors. The most precise data in the catalogue dates from the 1980s and has a precision of 60~ppm, but for much of it the formal errors are hundreds of ppm -- 350~ppm being a typical value.}. There is thus reason to doubt the reliability of the measurements of many of the nearer stars. Nevertheless the bifurcation trend is increasingly apparent beyond 50~pc. The LHB is not spherical, nor is the Sun at its centre, a fact which naturally explains the phenomenon. Here, in our much smaller sample, the two stars with higher polarizations have more northerly declinations than the rest of the sample. \citet{gontcharov17} places the shell of the LHB at 80--118~pc, a range that includes six of our seven targets. Therefore it is not surprising to find a sharp divergence in the magnitude of polarization amongst these stars. 

\citet{santos11} has gone into more detail, and plotted $p/d$ for a number of different regions inside, outside and within the ring associated with Loop I. It is worth consulting these maps (which can be found in \citealp{santos11}'s figures 7, 8 and 9) where our stars fall within the mapped regions. HD~165760 is just outside the ring in the direction of the North Polar Spur (NPS). In this region an increase to $\sim$600~ppm has been noted to occur between 130--140~pc. HD~165760 is considerably closer at a distance of 76~pc, yet has a similarly high polarization magnitude. The other higher polarization star in our sample, HD~216735 is not in a region mapped by \citet{santos11}. 

HD~138905 is within the loop I ring, in a region where polarization is seen to increase beginning within 50~pc. As such, the polarization of this star may be showing the effects of grain erosion from shocks associated with the evolution of Loop I. HD~17566 and HD~125473 lie near the edges of the Loop I ring but the \citet{santos11} $p/d$ plots do not reveal sufficient detail to be able to identify such a distinct rising polarization as for HD~138905. However, as previously discussed, past polarimetry was far less sensitive than later studies, so this does not preclude the possibility of the same processes affecting all three stars. \citet{sorrell95} modelled the destruction of Mie-scattering dust grains by shocks, based on the core-mantle grains of \citet{greenberg76,greenberg84}, finding that a shock speed of 55~km/s would produce $\lambda_{max}$ equal to 350~nm (for a polymeric grain sublimation energy of 0.05~eV). 

Examining the location of our target stars to the Na~I gas density maps of \citet{lallement03} (table \ref{tab:stars}) reveals that most of them are within or on the 20~m{\AA} equivalent width contour line. However, two of the three beyond that marker are those with high polarization. This supports the conclusion that they are more polarized because they are in the wall of the LHB, rather than within the cavity. This provides evidence for a correlation between gas and dust density. Such a correlation was briefly looked for by \citet{bailey10} without firm conclusions being drawn; we are in the process of a more comprehensive investigation. However, care should be exercised here, as Na~I is a recombinant species, so that much of the Na column density is missed if Na~II dominates \citep{sembach94}. This means that while we might use Na~I to trace the column density in the cooler walls of the LHB, this may not be reliable in warmer diffuse regions within the LHB. 

\subsection{The wavelength of maximum polarization in the local ISM}
\label{sec:colour}

Of particular interest here is that $\lambda_{max}$ of the low polarization targets is much shorter than that of the high polarization targets. The 2- and 3-parameter fits of the two high polarization targets, HD~165760 and HD~216735, give $\lambda_{max}$ as $\sim$550--557~nm, with errors less than 20~ppm. 550~nm is regarded as a typical value for ordinary stars beyond the LHB \citep{serkowski75}. Together with the higher polarization magnitude we conclude that these two stars are representative of the polarization within a wall of denser dust at the edge of the LHB. 

Here it is worth noting that the term ``walls'' doesn't reveal the full picture. Shocks from the multiple supernova and stellar winds that created the Local Bubble stir and mix the interstellar grains in the LHB. Large grains still associated with undisturbed or weakly disturbed dusty regions and shocked grains should both be found in the LHB \citep{frisch17}. Dusty regions with large grains tend to be found at the characteristic distances usually thought of as the walls, but there is no hard division between regions.

When considering the other stars the lower polarization magnitude, and associated uncertainty, requires we combine the data, which we do with the multi-parameter fits in figure \ref{fig:comb_Serk}. The parameter $K$ is a measure of the inverse width of the distribution, and since we seem to be probing mostly the long wavelength side of the curve, that width is not well fit, and we prefer the fit without $K$. The multi-target without $K$ fit gives $\lambda_{max}$ as 351.1~$\pm$~53.0~nm. Even though the formal error is quite large, it is clear this is significantly bluer than for the wall stars. A bluer $\lambda_{max}$ was also suggested by our earlier work \citep{marshall16} -- there the most probable value was determined to be 470~nm, but the optimum fit to the data was actually 315~nm. Such a blue value of $\lambda_{max}$ is sometimes referred to as ``super-Serkowski'' behaviour \citep{clayton95}, and since $K\propto\lambda_{max}$ the width of the peak in the Serkowski curve is broader than for more ordinary behaviour (\citealp{clayton95} also link a small $\lambda_{max}$ to greater near-UV extinction). As alluded to in the Introduction, a blue value for $\lambda_{max}$ is consistent with grains that have been shocked by past supernovae \citet{welty99}, as is believed to be the case for the LHB \citep{berghofer02, frisch17}, and is supported by UV spectroscopic measurements of nearby stars \citep{frisch11}.

The value of $\lambda_{max}$ is often used to draw conclusions about dust grain size. For instance, in the models of \citet{mathis86}, $\lambda_{max}$ is determined by the size-distribution of grains containing super-paramagnetic (SPM) particles (such as Fe). These SPM particles are thought to repel each other but coagulate with other materials, so form part of small grains when the grain size-distribution favours small grains, and become part of large grains with increased coagulation. In his adopted distribution the smallest size for particles contributing to alignment, and therefore interstellar polarization, is 50~nm. A key parameter related to the size-distribution of particles is $a'$, the radius of a grain having probability $e^{-1}$ of containing no SPM particles. \citet{mathis86} found that $a'$ and $\lambda_{max}$ were related by: \begin{equation}\label{eq:Mathis}a'=0.329\lambda_{max}^{2.17},\end{equation} where $\lambda_{max}$ is in $\mu$m. Which for $\lambda_{max}$ of 550 and 350~nm gives respectively $a'$ values of 90 and 34~nm. This, however, assumes the basic form of the size-distribution is unchanged\footnote{An assumption that leads to a larger value of $K$ for values of $\lambda_{max}$ smaller than 550~nm.}, which is unlikely to be the case in a medium recently shocked by supernovae.

As previously mentioned, \citet{sorrell95} made calculations for showing $\lambda_{max}$ is reduced in a medium shocked by a supernova. More recently \citet{slavin15} made calculations of grain destruction through shocks using more sophisticated shock models. They find that the final size distribution is influenced by the shock speed. For shock speeds of 50 to 200~km/s the relative abundance of the smallest grains is markedly increased. The effect is greatest for speeds of 100 to 150~km/s, at faster speeds the majority of grains of all sizes are blown out. \citet{slavin15}'s calculations suggest that with more extensive polarimetric data it may be possible to find the speeds of past shocks by determining $K$ more precisely.

Alternatively, in the scheme of \citet{papoular18} $\lambda_{max}$ values of 550~nm and 350~nm correspond to enstantite/fosterite ratios of $\sim$0.15 and 0.45 respectively.

\subsection{Deviation from interstellar-like behaviour}
\label{sec:intrinsic}

Classically interstellar polarization is characterised by a Serkowski curve, and consistency in $PA$ across wavelengths \citep{clarke10}. In terms of $PA$ consistency there is some scatter in the 425SP data compared to other bands for most objects. This could just be a result of lower instrumental precision in the bluest, lowest flux band. Night-to-night precision has been determined for HIPPI in a 500SP band \citep{bailey15} but not specifically for 425SP, so this is hard to gauge. However, a lower precision should not effect the mean accuracy given sufficient data. The 425SP $PA$ value is more often lower (more negative) than other bands than higher. So this could be an indication that the TP $u$ determination is greater than it should be. A 10~ppm shift in this value would reduce the 425SP $PA$ variance, with minimal impact on $p$. If, on the other hand, the 425SP scatter is a real reflection of the properties of the ISM, it could indicate the presence of multiple dust clouds made up of small particles with different alignments to the dominant clouds.

The HD~4150 data is clearly not described by a Serkowski curve -- which cannot have a minimum at optical wavelengths. Broadly, two possibilities exist to explain the HD~4150 observations. Either there is intrinsic polarization present along with interstellar polarization, or the measurements reflect multiple dust clouds with different properties along the line of sight. Though there is insufficient data available to rule this out\footnote{The best available maps would be those of \citet{berdyugin04, berdyugin14} except that they don't quite extend to this region.}, for two dust clouds to be responsible for the observed polarization with wavelength curve without any significant $PA$ rotation taking place, the two clouds would have to have very different $\lambda_{max}$ values -- one bluer than otherwise seen here, and one redder -- or have effectively opposite alignments and drastically different $K$ values. For HD~4150 the most likely explanation is that the polarization is intrinsic. HD~4150 is a binary system, with a K0~V star in a 9.8~yr orbit; this may be a factor, but as the flux of the secondary is only $\sim$1.6~\% of that of the primary component the companion would have to be far more polarized than other stars of its type have been found to be \citep{cotton17a}. As discussed in section \ref{sec:stars} entrained gas between binary stars can act as a scattering medium to produce polarization, but this polarization has little wavelength dependence, and the orbital period is too long to account for the measurement variability. The system also has a very slight infrared excess, detected photometrically in the far-IR. This excess is indicative of the presence of a debris disk, but the excess magnitude is too small to have an appreciable effect under ordinary circumstances. The combination of the companion and the scattering material causing the excess may be responsible in a similar fashion to that postulated in \citet{cotton16c}. If the star also has a near-IR excess this might have gone undetected, however current evidence associates such a property with nanoscale dust grains which do not appear to be strongly polarizing, but cannot be ruled out at the levels required here to explain the curve \citep{marshall16}. Another possibility is that the star is rapidly rotating (like Regulus as discussed in \citealp{cotton17b}) but highly inclined so as to have a low $v$sin$i$ value. In this case, an A0 star like HD~4150 would be expected produce polarization that flipped from being parallel to perpendicular to its rotation axis at bluer wavelengths than Regulus. Thus the shape of the curve in figure \ref{fig:3p2p_Serk} would describe a small intrinsic polarization oppositely aligned to a larger interstellar polarization.

\section{Conclusions}
\label{sec:conclusions}

We have made high precision polarimetric multi-band measurements of seven stars within or near the edge of the LHB. Subsequently we used this data to carry out full bandpass fitting of the Serkowski/-Wilking relation to determine the properties of interstellar polarization. One star, HD~4150, is not well fit by a Serkowski curve and so is probably intrinsically polarized. The most likely mechanism seems to be rapid rotation, if this is the case then the low $v$sin$i$ of the star indicates that it is highly inclined. 

The other six stars we find to be representative of interstellar polarization. Two stars, HD~165760 and HD~216735, have polarizations of $\sim$800~ppm and $\lambda_{max}$ $\sim$550~nm and we deduce they are in the wall of the LHB. The other four stars have much lower polarizations of $\sim$160~ppm or less and fit together have a $\lambda_{max}$ of $\sim$350~$\pm$~50~ppm. One of these stars, HD~138905, lies within the ring of Loop I at a distance where polarization is increasing sharply, and it is likely its polarization is the result of grains shocked by the expanding Loop I Superbubble. Another two stars, HD~17566 and HD~125473, lie near the edge of the Loop I ring and may be affected by the same process, or indeed by similar processes from other past supernova explosions within the LHB. Such a blue determination for $\lambda_{max}$ supports the notion that the larger particles in this region have either been shocked or swept up by past supernovae leaving the region to be defined by smaller grains. The data for the final star, HD~206453, has greater uncertainties and may be fit by a range of $\lambda_{max}$ values, it lies further away from Loop I than the other three stars.

For studies of stars in nearby space where a wavelength dependence of interstellar polarization needs to be assumed, we recommend using the formulation of \citet{whittet92} with values of $\lambda_{max}$ of 550~nm and 350~nm for respectively, stars near the edge of the LHB with higher polarizations, and stars within it with lower polarizations. However, we caution that in the latter case this may only apply to stars near to or within the Loop I ring. Clearly, larger samples are needed to establish the robustness of the trends inferred here. More work needs to be done to assess other regions, and where this is not possible, a cautions approach might be to take an intermediate value, such as 470~nm as found most likely by \citet{marshall16}.

\section*{Acknowledgements}

JPM acknowledges research support by the Ministry of Science and Technology of Taiwan under grants MOST104-2628-M-001-004-MY3 and MOST107-2119-M-001-031-MY3, and Academia Sinica grant AS-IA-106-M03. This research has made use of the SIMBAD database, operated at CDS, Strasbourg, France. The authors thank the Director and staff of the Australian Astronomical Observatory for their advice and support with interfacing HIPPI to the AAT and during the two observing runs on the telescope. Further, we would like to thank Behrooz Karamiqucham, Dag Evensberget and Jinglin Zhao for their assistance in acquiring the data, and Christoph Bergmann for useful discussions. We thank the anonymous referee for their useful feedback.

\small\bibliographystyle{mnras}
\bibliography{ism_colour2} 
\bsp

\appendix

\label{lastpage}

\end{document}